\documentclass{ws-procs9x6}

\usepackage{url}
\newcommand{\ds}{{\sffamily DarkSUSY}}

\newcommand{\cpc}[3]{Comm.\ Phys.\ Comm.\ {\bf #1} (#2) #3}
\newcommand{\prl}[3]{Phys.\ Rev.\ Lett. {\bf #1} (#2) #3}

\newcommand{\prd}[3]{Phys.\ Rev.\ {\bf D#1} (#2) #3}

\newcommand{\href}[2]{#1}

\begin{document}

\title{DarkSUSY -- A numerical package for supersymmetric dark matter
calculations\footnote{\uppercase{P}reprint: \uppercase{CWRU-P14-02}}}

\author{
Paolo Gondolo$^{\mbox{\lowercase{a}}}$\footnote{\uppercase{E}-mail: pxg26@po.cwru.edu}~,
\underline{Joakim Edsj\"o}$^{\mbox{\lowercase{b}}}$\footnote{\uppercase{R}apporteur, \uppercase{E}-mail: edsjo@physto.se}~,
Piero Ullio$^{\mbox{\lowercase{c}}}$\footnote{\uppercase{E}-mail: ullio@he.sissa.it}~,
Lars Bergstr\"om$^{\mbox{\lowercase{b}}}$\footnote{\uppercase{E}-mail: lbe@physto.se}~,
Mia Schelke$^{\mbox{\lowercase{b}}}$\footnote{\uppercase{E}-mail: schelke@physto.se}~
and Edward A. Baltz$^{\mbox{\lowercase{d}}}$\footnote{\uppercase{E}-mail: eabaltz@physics.columbia.edu}}

\address{a) Department of Physics, Case Western Reserve University, 
10900 Euclid Ave., Cleveland OH 44106-7079, USA\\
b) Department of Physics, Stockholm University\\
AlbaNova, SE-106 91 Stockholm, Sweden\\
c) SISSA, via Beirut 4, I-34014 Trieste, Italy\\
d) ISCAP
Columbia Astrophysics Laboratory, 550 W 120th St., Mail Code 5247
New York, NY 10027, USA}


\maketitle

\abstracts{ The question of the nature of dark matter in the Universe remains
  one of the most outstanding unsolved problems in basic science.  One of the
  best motivated particle physics candidates is the lightest supersymmetric
  particle, assumed to be the lightest neutralino.  We here describe \ds, an
  advanced numerical {\sc FORTRAN} package for supersymmetric dark matter
  calculations.  With \ds\ one can: (i) compute masses and compositions of
  various supersymmetric particles; (ii) compute the relic density of the
  lightest neutralino, using accurate methods which include the effects of
  resonances, pair production thresholds and coannihilations; (iii) check
  accelerator bounds to identify allowed supersymmetric models; and (iv) obtain
  neutralino detection rates for a variety of detection methods, including
  direct detection and indirect detection through antiprotons, gamma-rays and
  positrons from the Galactic halo or neutrinos from the center of the Earth or
  the Sun.  }

\section{Introduction}

One of the favourite candidates for the dark matter is a Weakly Interacting
Massive Particle, a WIMP. In supersymmetric extensions of the standard model,
the neutralino emerges as a natural WIMP candidate for the dark matter of the
universe.

Over several years, we have developed analytical and numerical tools for
dealing with the sometimes complex calculations necessary to go from given
supersymmetric input parameters to actual quantitative predictions of the
neutralino relic density in the Universe, and of the direct and indirect
detection rates of neutralino dark matter.  In 2000, we made available the
first public release of \ds\ \cite{idm2000-ds}. Here we outline the different
components of \ds, and comment on the updates and improvements in the coming
release.
The latest version of \ds\ can be downloaded from the official \ds\ website,
{\tt http://www.physto.se/\~{}edsjo/darksusy/}.

\section{Supersymmetric model in \ds}

\ds\ implements the general structure of the Minimal Supersymmetric Standard
Model (MSSM), with R-parity and CP conservation (except in the quarks CKM
matrix). The parameter space of the MSSM is specified by 124 a priori free
parameters\cite{dimo}.

Most of the \ds\ code does not make any specific assumption about the 124
parameters (except for R-parity and CP conservation, as mentioned). However, in
the version of \ds\ soon to be released, the supersymmetric mass spectrum and
the particle mixing matrices are computed under one of two classes of
restrictions on the choice of supersymmetric parameters: (1) seven input
parameters at the electroweak scale, as was available in the previous release
(see \cite{ds,bg} for details), or (2) five parameters at the GUT scale, in the
context of minimal supergravity (mSUGRA) implemented via an interface to
ISASUGRA\cite{isajet}. Beyond these two possibilities, the user can in
principle implement his own flavor of MSSM.

\ds\ calculates all masses, mixings, and most of the vertices entering the
Feynman rules. All of these are available to the user. \ds\ includes several
options for the loop corrections to the Higgs
masses\cite{feynhiggs1,drees92,carena95,carena96}, and for the loop corrections
to the neutralino and chargino masses\cite{NeuLoop1,NeuLoop2}.

\section{Accelerator bounds}

Accelerator bounds can be checked by a call to a subroutine.  By
modifying an option, the user can impose bounds as of different
moments in time.  The default option in the coming public release adopts the
2002 limits by the Particle Data Group\cite{PDG}.
The user is also free to use his own routine to
check for experimental bounds, in which case he or she would only need
to provide an interface to \ds.

\section{Calculation of the relic density}

\ds\ calculates the neutralino relic density by fully computing the thermal
average of the effective neutralino annihilation cross section, including all
resonances, thresholds and coannihilations, and then solves the Boltzmann
equation numerically with the methods given in \cite{GondoloGelmini,coann}.

A major update in the next public release is the inclusion of coannihilations
with squarks and sleptons. Coannihilations are important in the computation of
the relic density when other particles the neutralino can convert to are close
in mass to the lightest neutralino. These other particles are abundant
at the time of neutralino freeze-out in the early Universe, and their
annihilation rate must be included in the calculation. The current public
release of \ds\ includes coannihilation processes between all neutralinos and
charginos; the next release includes coannihilations between all the above and
also squarks and sleptons.

\ds\ includes all two-body final states that occur at tree level, and
neutralino annihilation into $gg$, $\gamma \gamma$ and $Z\gamma$ that occur at
the 1-loop level.

\section{Detection rates}

The different detection rates for neutralino dark matter have been
calculated by many authors in the past.  We will here only give a
brief review about what is included in \ds, and which calculations
they are based on.  For a more extensive list of references, we refer
to \cite{ds}.

\subsection{Halo models}

Currently implemented in \ds\ is the spherical family of halo 
profiles 
  $\rho(r) \propto 1/\left((r/a)^{\gamma}\;[1+(r/a)^{\alpha}]^
  {(\beta-\gamma)/\alpha}\right)$
where e.g.\ the Navarro, Frenk and White profile\cite{navarro} is
given by $(\alpha,\beta,\gamma)=(1,3,1)$ and the isothermal sphere is
given by $(\alpha,\beta,\gamma)=(2,2,0)$.  The velocity distribution
is assumed to be a standard isotropic gaussian distribution.

\subsection{Direct detection}\label{subs:direct}

These routines calculate the spin-dependent and spin-independent 
scattering cross sections on protons and neutrons assuming the quark 
contributions to the nucleon spin from \cite{SMC}. The older set of 
data from \cite{jaffe} is also available as an option, and the user 
can set his/her own values if desired.

\subsection{Monte Carlo simulations}

In several of the indirect detection processes described below we need the
yield of different particles per neutralino annihilation.  The hadronization
and/or decay of the annihilation products have been simulated with {\sc
  Pythia}\cite{pythia} and the results have been tabulated.  These tables are
incorporated in \ds.

\subsection{Neutrinos from the Sun and Earth}

Neutralinos can accumulate in the Earth and the Sun where they can 
annihilate pair-wise producing high energy muon neutrinos. The 
branching ratios for different annihilation channels are calculated 
and the {\sc Pythia} simulations are used to evaluate the yield of 
neutrinos. Neutrino interactions in the Sun as well as the charged 
current neutrino-nucleon interaction near the detector are also 
simulated with {\sc Pythia}.

There are routines to calculate a) the neutrino flux, b) the neutrino-to-muon
conversion rate and c) the neutrino-induced muon flux either differential in
energy and angle or integrated within an angular cone and above a given
threshold.  The new population of neutralinos in the solar system described in
\cite{dk1,dk2,dkpop} (arising from neutralinos that have scattered in the
outskirts of the Sun) can optionally be included as well.

\subsection{Antiprotons from halo annihilations}

Neutralinos can also annihilate in the Milky Way halo producing e.g.\
antiprotons.  These propagate in the galaxy before reaching us.  We
have implemented the propagation method described in \cite{pbar}. 
Optionally, the antiproton fluxes can also be solar modulated with the
spherically symmetric model of \cite{GleesonAxford}.  There are also
other propagation models\cite{pbarother} available as options.  The
antiproton fluxes are given differential in energy.

\subsection{Positrons from halo annihilations}

Neutralino annihilations in the halo can also produce positrons. The 
flux of positrons is calculated with the propagation model in 
\cite{baltz} (with two choices of the energy dependence of the 
diffusion constant). The models in \cite{kamturner} or \cite{moskstrong} 
can also be used as an option. The positron fluxes are given differential in energy.

\subsection{Gamma rays from halo annihilations}

Halo annihilations can also produce gamma rays.  These are either monochromatic
(produced from 1-loop annihilation\cite{lp} into $\gamma\gamma$ and
$Z\gamma$\cite{ub}) or have a continuous energy spectrum (produced from
$\pi^{0}$ decays in quark jets\cite{beu}).

The flux of gamma rays can be obtained in any given direction on the
sky for the user's choice of $(\alpha, \beta, \gamma)$ in the halo
profile.  There are also routines to average the flux over a chosen
angular resolution.  The continuous gamma rays use {\sc Pythia}
simulations to calculate the gamma ray flux (differential in energy,
or integrated above an energy threshold).

\subsection{Neutrinos from halo annihilations}

Neutralino annihilations in the halo can also produce neutrinos. Although the
neutrino fluxes are small, \ds\ contains routines to calculate (a) the neutrino
flux, (b) the neutrino-to-muon conversion rate, and (c) the neutrino-induced
muon flux either differential in energy and angle or integrated within an
angular cone and above a given energy threshold.

\section{Conclusions}

Over the years we have developed a numerical package, \ds, for neutralino dark
matter calculations in the Minimal Supersymmetric Standard Model, MSSM.  Since
the initial public release in 2000, the main improvements to the code are the
inclusion of mSUGRA models via an interface to ISASUGRA and the inclusion of
all squark and slepton coannihilations.  These additions will be available in
the next public release.

We have here presented an overview of what the program can do, and refer the
reader to the upcoming paper \cite{ds} and manual \cite{ds-manual}, where the
details will be given. A test program, provided with the distribution, shows in
more detail how \ds\ is used.

\section*{Acknowledgments}
L.B.\ and J.E.\ thank the Swedish Research Council for their 
support.

\end{document}